\documentclass[prd,twocolumn,showpacs,preprintnumbers,superscriptaddress,floatfix,nofootinbib]{revtex4-1}
\usepackage{subfigure}
\usepackage{amsfonts}
\usepackage{enumitem}
\usepackage{graphicx,,booktabs}
\usepackage{amssymb}
\usepackage{amsmath,txfonts,ulem}
\usepackage{graphicx}
\usepackage{dcolumn}
\usepackage{multirow}
\usepackage{color}
\usepackage{bm}
\usepackage{ulem}
\usepackage{makecell}
\usepackage[colorlinks,
citecolor=blue,
anchorcolor=green,
menucolor=orange,
linkcolor=red,
filecolor=red,
runcolor=pink,
urlcolor=blue,
frenchlinks=red]{hyperref}

\allowdisplaybreaks[4]

\begin{document}

\title{First characterization of the scattering length distribution of
vector meson interaction with deuteron}
\author{Xiao-Yun Wang}
\email{xywang@lut.edu.cn}
\affiliation{Department of physics, Lanzhou University of Technology,
Lanzhou 730050, China}
\affiliation{Lanzhou Center for Theoretical Physics, Key Laboratory of Theoretical Physics of Gansu Province, Lanzhou University, Lanzhou, Gansu 730000, China}
\author{Chen Dong}
\affiliation{Department of physics, Lanzhou University of Technology,
Lanzhou 730050, China}
\author{Quanjin Wang}
\affiliation{Department of physics, Lanzhou University of Technology,
Lanzhou 730050, China}

\begin{abstract}
Under the framework of the vector meson dominance model (VMD), the absolute
scattering lengths of light vector mesons ($\phi$, $\omega$, and $\rho$)
with the deuteron are calculated from the cross sections of vector meson
photoproduction off a deuteron. Additionally, a fitting function is used to
predict the scattering lengths of the heavy vector mesons $J/\psi$-$d$ and $%
\Upsilon$-$d$. Based on these results, the distribution of the scattering
lengths $|\alpha_{V d}|$ between vector mesons and the deuteron is presented
for the first time and compared with the scattering lengths $|\alpha_{V p}|$
between vector mesons and the proton. It is found that as the mass of the
vector meson increases, the scattering lengths $|\alpha_{V d}|$ become
closer to $|\alpha_{V p}|$.
In addition, the correlation analysis indicates a strong positive relationship between the scattering length and the vector meson radius, with a high correlation coefficient.
The implications of results are essential for improving our understanding of the interaction between hadron and nucleus.

\end{abstract}

\maketitle


\section{Introduction}

The photoproduction of electrically neutral vector mesons is a strong
diffraction process \cite{Aachen-Bonn-Hamburg-Heidelberg-Munich:1974olt},
which is similar to elastic scattering due to the same quantum number of
incoming and outgoing particles. The process has been fully interpreted
through the vector meson dominated model (VMD) \cite{Bauer:1977iq}, which
suggests that photons fluctuate into virtual light vector mesons and then
scatter elastically through the target. The VMD model is not limited to
investigating the photoproduction of vector mesons from protons, but is also
commonly used to explore the photoproduction of vector mesons off a deuteron
\cite{VMD,Rogers:2006ug,Frankfurt:1997ss}.

However, when the photon energy is close to the production threshold, the
pseudoscalar meson exchange of the $t$-channel can also produce
corresponding contributions \cite{Oh:2000pr}, which complicates the
investigation of photoproduction dynamics of vector mesons on a proton
target. Additionally, the complexity at the threshold may result from $s$%
-channel nuclear resonance. Therefore, the photoproduction of vector mesons
off a deuteron can avoid this complexity \cite{CLAS:2018avi}. $\rho$ and $%
\omega$ are the most common light vector mesons with a component quark mass
of $m_u=m_d=330$ MeV \cite{Zhao:1998rt}. Investigating the reaction of
deuterons to heavy vector mesons through the scattering process of $\rho$
and $\omega$-deuteron interaction can be enlightening.

The differential cross section of $\gamma d$$\rightarrow$$\rho d$ channel
was first measured by R. L. Anderson et al. in 1971 \cite{Anderson:1971ar}.
They measured the differential cross section of $\rho$ photoproduction off a
deuteron at incident photon energies of $6$, $12$ and $18$ GeV using the
SLAC $1.6$ GeV/c spectrometer. Since $\rho$-$d$ can be treated as an elastic
scattering process, the authors used the Glauber theory \cite%
{Franco:1965wi,Harrington:1968sj} which can well describe $\pi$-$d$
scattering \cite{Fellinger:1969cw,Bradamante:1970fh} to study $\rho$-$d$.
The result showed that the Glauber theory can well describe the cross
section of $\gamma d$$\rightarrow$$\rho d$ channel in the small $t$ region.
In 1974, P. Benz et al. \cite{Aachen-Bonn-Hamburg-Heidelberg-Munich:1974olt}
measured the cross section of $\gamma d$$\rightarrow$$\rho d$ and $\gamma d$$%
\rightarrow$$\omega d$ in the incident photon energy range of $1.00 \sim 5.00
$ GeV at DESY and presented the effective mass distribution.

In 1997, the VMD model was used to study the photoproduction of vector
mesons at photon energies of $3\sim30$GeV and to explore the space-time
evolution of small-size quark-gluon configurations \cite{Frankfurt:1997ss}.
This model predicted the differential cross sections of $\rho$
photoproduction and electroproduction off a deuteron at different energies.
More recently, in 2018, the CLAS at Jefferson Lab \cite{CLAS:2018avi}
measured a new set of $\omega$ photoproduction data off a deuteron at $%
E\in(1.4, 3.4)$ GeV. Then in 2019, Ref. \cite{Chetry:2019rpw} analyzed this
data in detail and presented a new set of $\rho$-$d$ differential cross
sections at $E\in(1.4, 3.4)$ GeV. These experimental data provide a
prerequisite for further investigation of the vector meson-deuteron ($V$-$d$%
) interaction.

The absolute value of the scattering length characterizes the strength of
the $V$-$d$ interaction. The scattering length of vector meson-proton $%
|\alpha_{V p}|$ interactions has been extensively studied using VMD models
\cite%
{Wang:2022zwz,Strakovsky:2014wja,Strakovsky:2020uqs,Titov:2007xb,Wang:2022tuc,Pentchev:2020kao,Wang:2022xpw,Strakovsky:2021vyk}%
. In particular, it has been shown that the scattering length of $V$-$p$
exhibits an inverse relationship with the mass (excluding $\rho$-$p$), which
is $|\alpha_{V p}|$$\propto$$\exp{(\frac{1}{M_V})}$. In our previous work \cite%
{Wang:2022tuc}, we extracted the scattering length of $\phi$-$d$ ($%
|\alpha_{\phi d}|$) to be $0.014 \pm 0.002$ fm, which is about seven times
smaller than $|\alpha_{\phi p}|$. To further observe the behavior of
scattering lengths of different vector meson and deuteron interactions, we
aim to calculate the scattering length of more vector meson-deuteron
interactions using the VMD model.

In this study, we focus on calculating the scattering lengths of $\omega$
and $\rho$ mesons interacting with the deuteron. To achieve this, we utilize
the considerable photoproduction cross sections of $\rho$-$d$ \cite%
{Chetry:2019rpw} and $\omega$-$d$ \cite{CLAS:2018avi} that have been
accumulated through experiments. Furthermore, we also predict the scattering
length of heavy vector mesons ($J/\psi$, $\Upsilon$) interacting with the
deuteron. We discuss the implications of our results based on the size of
hadrons and the extent to which they are embedded in the nucleus.

To provide a clear outline of our work, we first introduce the VMD model and
explain how it is used to associate the differential cross section of vector
meson with the scattering length in Section \ref{sec2}. Subsequently, we
calculate the scattering lengths of $\omega$-$d$ and $\rho$-$d$ and present
our predictions for the $J/\psi$-$d$ and $\Upsilon$-$d$ scattering lengths.
Finally, we summarize our results briefly in Section \ref{sec3}.

\section{The scattering length for light vector meson-deuteron interaction}

\label{sec2}  The scattering of photons and deuterons is an essential
photoproduction process of vector mesons ($\omega$, $\rho$ and $\phi$). The
VMD model \cite{VMD,Rogers:2006ug,Frankfurt:1997ss} provides a unique method
to investigate the $\gamma d$$\rightarrow$$V d$ channel, where $d$
represents the deuteron. In the Ref. \cite{Titov:2007xb}, the VMD model
connects the differential cross-section of $\gamma p$$\rightarrow$$V p$ to
the scattering length of vector meson-proton ($V$-$p$) interaction. Given
the properties of the VMD model, which also can be used to relate the $V$-$d$
interaction scattering length $|\alpha_{V d}|$ with the differential cross
section $d \sigma / d t$ of vector meson from the deuteron at the
near-threshold \cite{Titov:2007xb},
\begin{equation}
\left.\frac{d \sigma}{d t}\right|_{t=t_{thr} }=\frac{\pi^2\alpha_{em}}{%
g_V^2\left|\mathbf{{p}_1}\right|^2} \cdot\left|\alpha_{V d}\right|^2,
\label{eq:1}
\end{equation}
where $\alpha_{em}$ is the fine coupling constant, $t_{thr}$ is the
production threshold of vector meson $t_{thr}=-M_V^2 m_d/(M_V+m_d)$ ($m_d$
is the mass of the deuteron) and $g_V$ is the VMD coupling constant,
\begin{equation}
g_V = \sqrt{\frac{\pi \alpha_{em}^2M_V}{3 \Gamma_{e^+e^-}}}.
\end{equation}
$\Gamma_{e^+e^-}$ is the lepton decay width and $M_V$ is the meson mass. In
the Eq. (\ref{eq:1}), $|\mathbf{{p}_1|}$ is the initial momentum in the
center of mass frame,
\begin{equation}
	\left|\mathbf{p}_1\right|=\frac{1}{2 W} \sqrt{W^4-2\left(m_1^2+m_2^2\right) W^2+\left(m_1^2-m_2^2\right)^2},
\end{equation}
and $|\mathbf{{p}_3|}$ is the final momentum,
\begin{equation}
	\left|\mathbf{p}_3\right|=\frac{1}{2 W} \sqrt{W^4-2\left(m_3^2+m_4^2\right) W^2+\left(m_3^2-m_4^2\right)^2},
\end{equation}
where $W$ is the center of mass energy.

We assume that $R=|\mathbf{{p}_3|/|{p}_1|}$, so the differential cross
section can be written as a function of $R$, and the relation between
scattering length $|\alpha_{V d}|$ and differential cross section $d \sigma
/d t (R)$ can be rewritten as,
\begin{equation}
|\alpha_{Vd}| = \frac{|\mathbf{{p}_3|}g_V}{R\pi} \sqrt{\frac{1}{\alpha_{em}}%
\left.\frac{d \sigma}{d t}\right|_{t=t_{t h r}}}.  \label{eq:5}
\end{equation}

\begin{figure}[htbp]
	\begin{center}
		\includegraphics[scale=0.40]{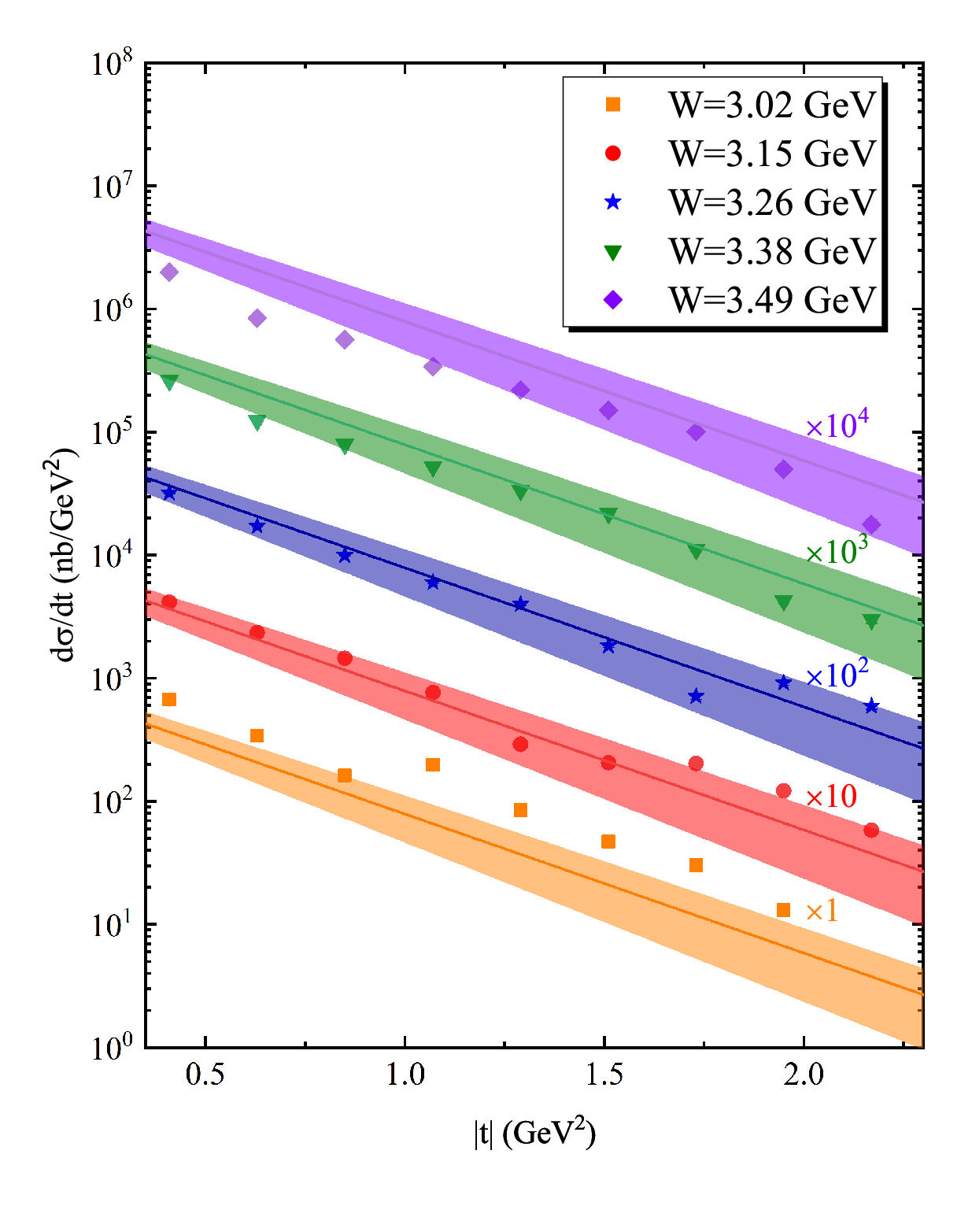}
	\end{center}
	\caption{The fitting results between the exponential function $d\protect%
		\sigma /d t$ and the experimental differential data \protect\cite%
		{Chetry:2019rpw} of $\protect\rho$ photoproduction off a deuteron. The error band is the error of slope $b$, the $\chi^2$/d.o.f = $2.74$.}
	\label{fig:fit}
\end{figure}

For $\rho$ mesons, the differential cross sections of $\gamma d$$\rightarrow$%
$\rho d$ at $E\in[1.50,2.30]$ GeV ($W\in[3.02,3.49]$ GeV) involved in the Ref.
\cite{Chetry:2019rpw} are selected to extract the scattering length $%
|\alpha_{\rho d}|$. The reason for choosing the interval is that it is
closest to the $\rho$ production threshold $W_{thr}=2.646$ GeV in the
available data.  Here, the experimental data are fitted through a
differential form of an exponential function $d \sigma /dt = A \exp{%
(b(t-t_{thr}))}$ \cite{Pentchev:2020kao} to obtain the $d \sigma
/dt|_{t=t_{thr}}$, where $A$ and $b$ are two free parameters.
The cross section of $\rho$ photoproduction on a deuteron is analyzed through a joint fit, resulting in values of $A=0.008\pm0.003$ and $b=2.60\pm0.38$ GeV$^{-2}$, the value of $\chi^2$/d.o.f is calculated as $2.74$. The relevant fitting results are presented in Fig. \ref{fig:fit}, and the corresponding
the scattering length of $\rho$-d are shown in Tab. \ref{tab:table1}.

\begin{table}[htbp]
\caption{The $R$, and $|\protect\alpha_{\protect\rho d}|$ for different center of mass energies $W$.}
\label{tab:table1}{\small \resizebox{\linewidth}{!}{
			\begin{tabular}{ccccc}
				\hline
				\hline
				W (GeV)  & $3.02$&$3.15$ & $3.26$& $3.38$  \\
				\hline
				R & $0.73$ & $0.79$ & $0.82$ & $0.85$\\
				\hline
				$|\alpha_{\rho d}|$ (fm)& $0.051\pm 0.007$ & $0.056\pm0.008$ & $0.060\pm0.009$ & $0.064 \pm 0.009$\\
				\hline
				\hline
				W (GeV)&$ 3.49$&-&-&-\\
				\hline
				R & $0.87$&-&-&-\\
				\hline
				$|\alpha_{\rho d}|$ (fm)& $0.068\pm0.010$&-&-&-\\
				\hline
				\hline
			\end{tabular}
		}  }
\end{table}

\begin{figure}[htbp]
\begin{center}
\includegraphics[scale=0.40]{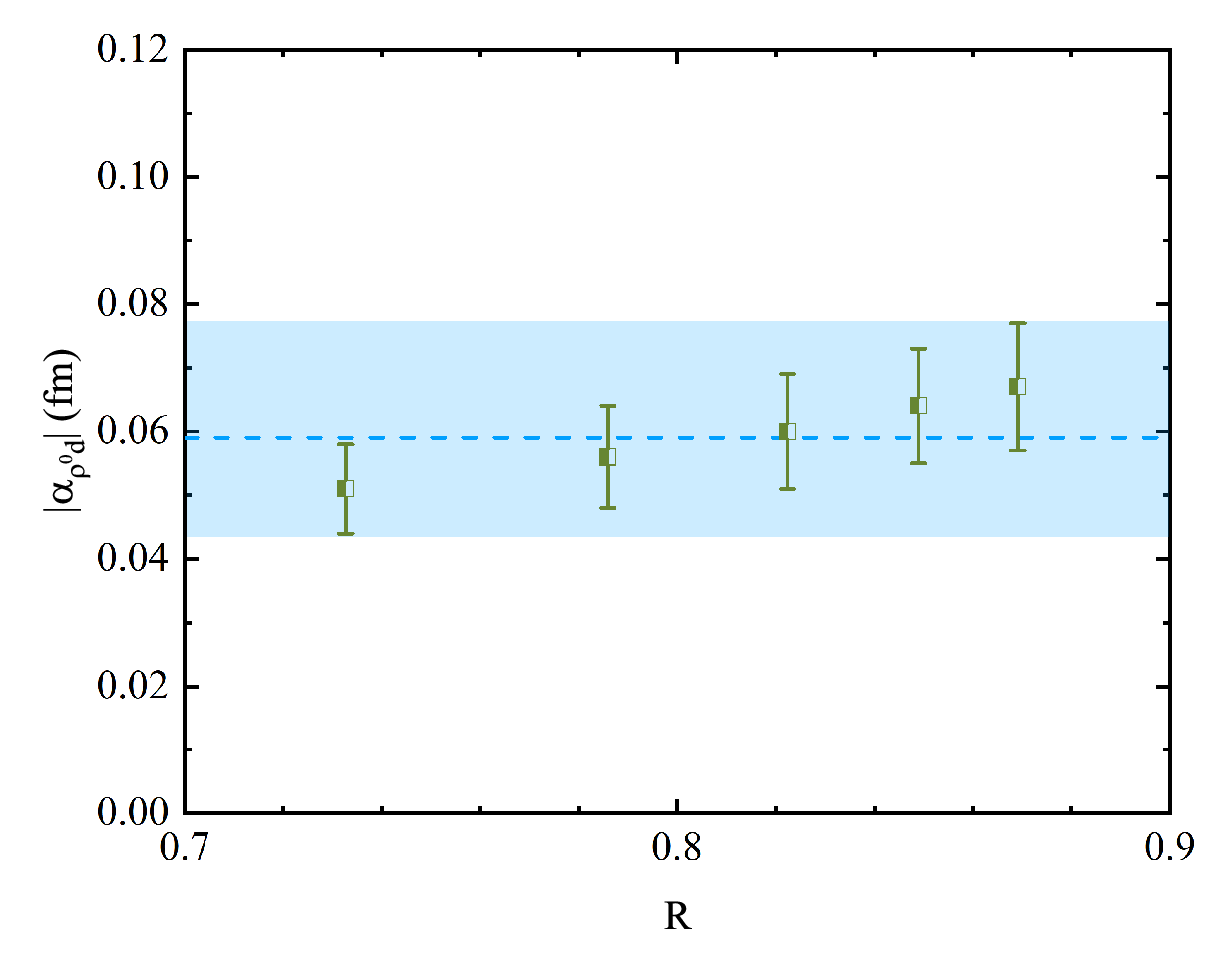}
\end{center}
\caption{The $|\protect\alpha_{\protect\rho d}|$ derived at different $R$.
The cyan dashed line is the RMS value $\protect\sqrt{\left\langle |\protect%
\alpha^{2}_{\protect\rho d}|\right\rangle}=0.060 \pm 0.017$ fm. The
differential cross section data of $\protect\rho$ photoproduction is from
Ref. \protect\cite{Chetry:2019rpw}. The cyan error band is the error band of
$|\protect\alpha_{\protect\rho d}|$.}
\label{fig:rho-d}
\end{figure}

The $|\alpha_{\rho d}|$ at different $R$ is shown in Fig. \ref{fig:rho-d}
(the green squares),  indicating a rising trend as $R$ increases.
In order to obtain an average calculated value, the root mean square
(RMS) is calculated. The  cyan dashed line shows the RMS of these five
values, which is calculated as $\sqrt{\left\langle |\alpha^{2}_{\rho
d}|\right\rangle}=0.060 \pm 0.017$ fm. The error band is the error that
takes into account all the results.

The differential cross sections of $\gamma d$$\rightarrow$$\omega d$ at $E\in%
[1.6,3.1]$ GeV ($W\in[3.09,3.89]$ GeV) measured by CLAS at JLAB \cite%
{CLAS:2018avi} can be used to extract $|\alpha_{\omega d}|$.
The empirical formula $d \sigma /dt = A \exp{%
	(b(t-t_{thr}))}$ is also used to make a joint fit to the experimental data of $\omega$, and $A=4.77\pm1.64 \times 10^{-4}$ and $b=2.69\pm0.36$ GeV$^{-2}$ are obtained, the $\chi^2$/d.o.f is $0.44$.
The variation of $|\alpha_{\omega d}|$ with $R$ extracted from the CLAS data \cite%
{CLAS:2018avi} is shown in Fig. \ref{fig:omega-d} (the olive green diamond),
and the relevant calculation results are listed in Tab. \ref{tab:table2}. Furthermore, one also calculates its RMS as $\sqrt{%
\left\langle |\alpha^{2}_{\omega d}|\right\rangle}=0.052 \pm 0.019$ fm,
which is represented by the purple double dot dash line in Fig. \ref%
{fig:omega-d}.

\begin{figure}[htbp]
	\begin{center}
		\includegraphics[scale=0.40]{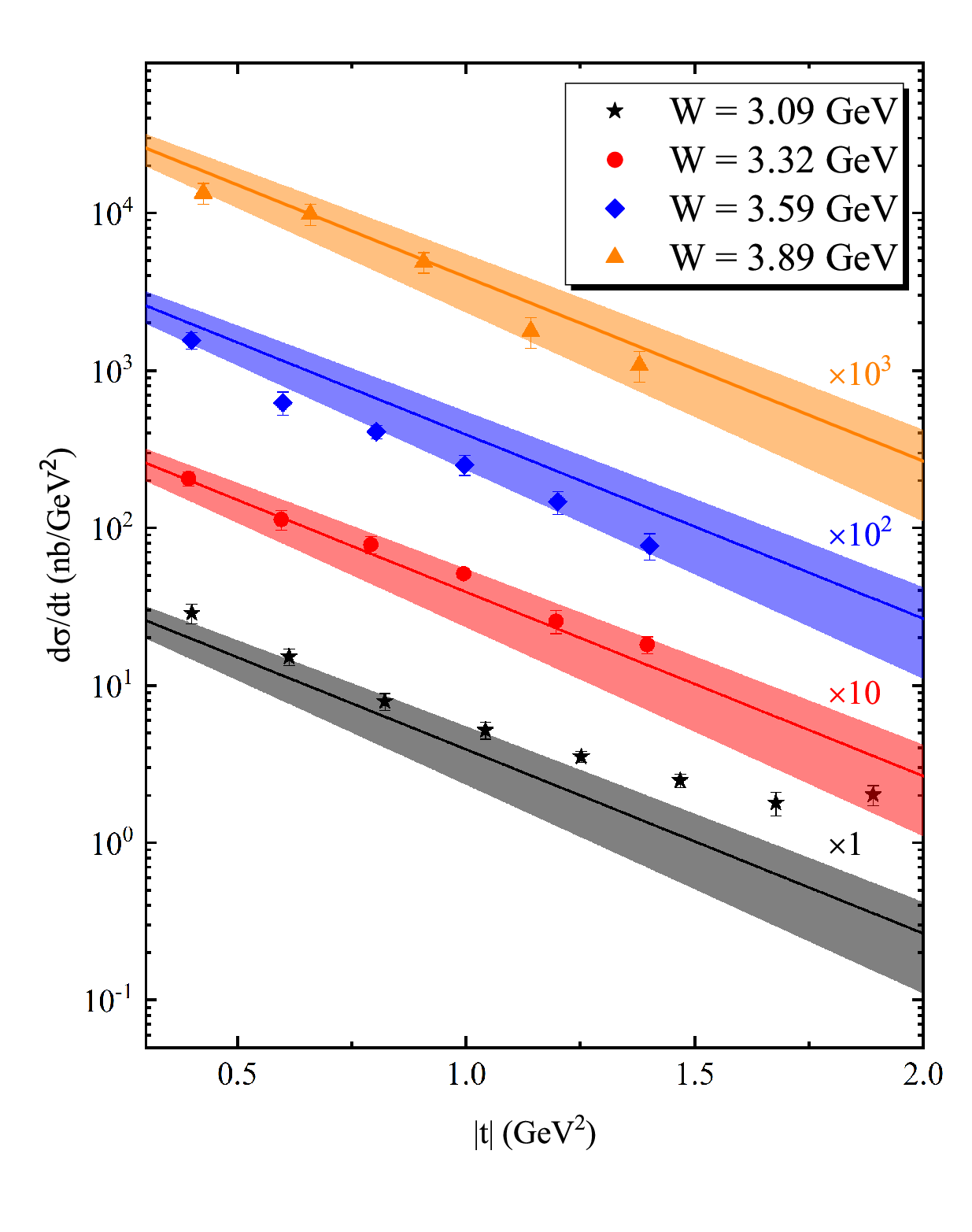}
	\end{center}
	\caption{The fitting results between the exponential function $d\protect%
		\sigma /d t$ and the experimental differential data \cite{CLAS:2018avi} of $\protect\omega$ photoproduction off a deuteron. The error band is the error of slope $b$, the $\chi^2$/d.o.f = $0.44$.}
	\label{fig:omega-d}
\end{figure}

\begin{figure}[htbp]
\begin{center}
\includegraphics[scale=0.40]{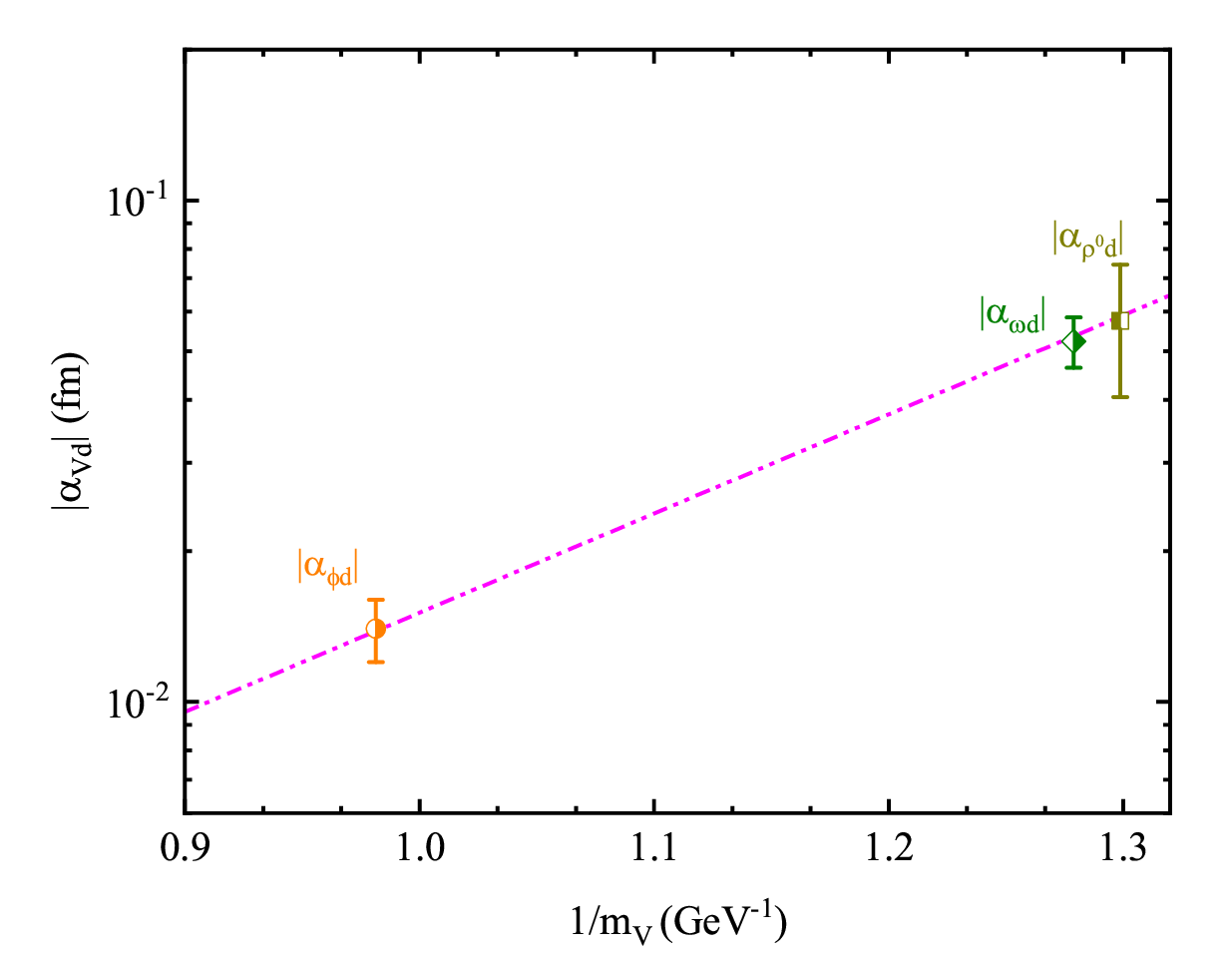}
\end{center}
\caption{The scattering length of vector mesons and deuteron interaction
derived by VMD model. The orange circle is the scattering length of $\protect%
\phi$-$d$ taken from our previous work \protect\cite{Wang:2022tuc}. The
olive green diamond and the dark yellow square are the result of $|\protect%
\alpha_{\protect\omega d}|$ and $|\protect\alpha_{\protect\rho d}|$,
respectively.}
\label{fig:v-d}
\end{figure}

At present, we have calculated the absolute value of the scattering lengths
of $\rho$-$d$, $\omega$-$d$ and $\phi$-$d$.  In order to explore the
relation between the mass of vector meson and $|\alpha_{V d}|$, a fitting
function is obtained by the above results,
\begin{equation}
|\alpha_{V d}|=\exp{[-8.73+(4.54\pm0.30) \frac{1}{M_V}]}.  \label{eq:6}
\end{equation}
As shown in Fig. \ref{fig:v-d}, it is found that the scattering lengths $%
|\alpha_{V d}|$ decreases as the mass of the vector meson increases.
However, it should be mentioned that we did not take the value of $%
|\alpha_{\rho d}|$ into account when calculating the fitting function above.
There are two main reasons, one is because $|\alpha_{\rho d}|$ itself has a
large error. Another reason is that the width of $\rho$ is much larger than
that of other vector mesons \cite{ParticleDataGroup:2022pth}, which also
leads to inaccuracy in our calculation of $\rho$-$N$ scattering length using
VMD model \cite{Wang:2022zwz}. It is only a coincidence that the scattering
lengths of $\rho$-$d$ and $\omega$-$d$ are close at present, and more
relevant discussions and modifications to the VMD model will be continued in
the future work.

\begin{figure}[htbp]
\begin{center}
\includegraphics[scale=0.40]{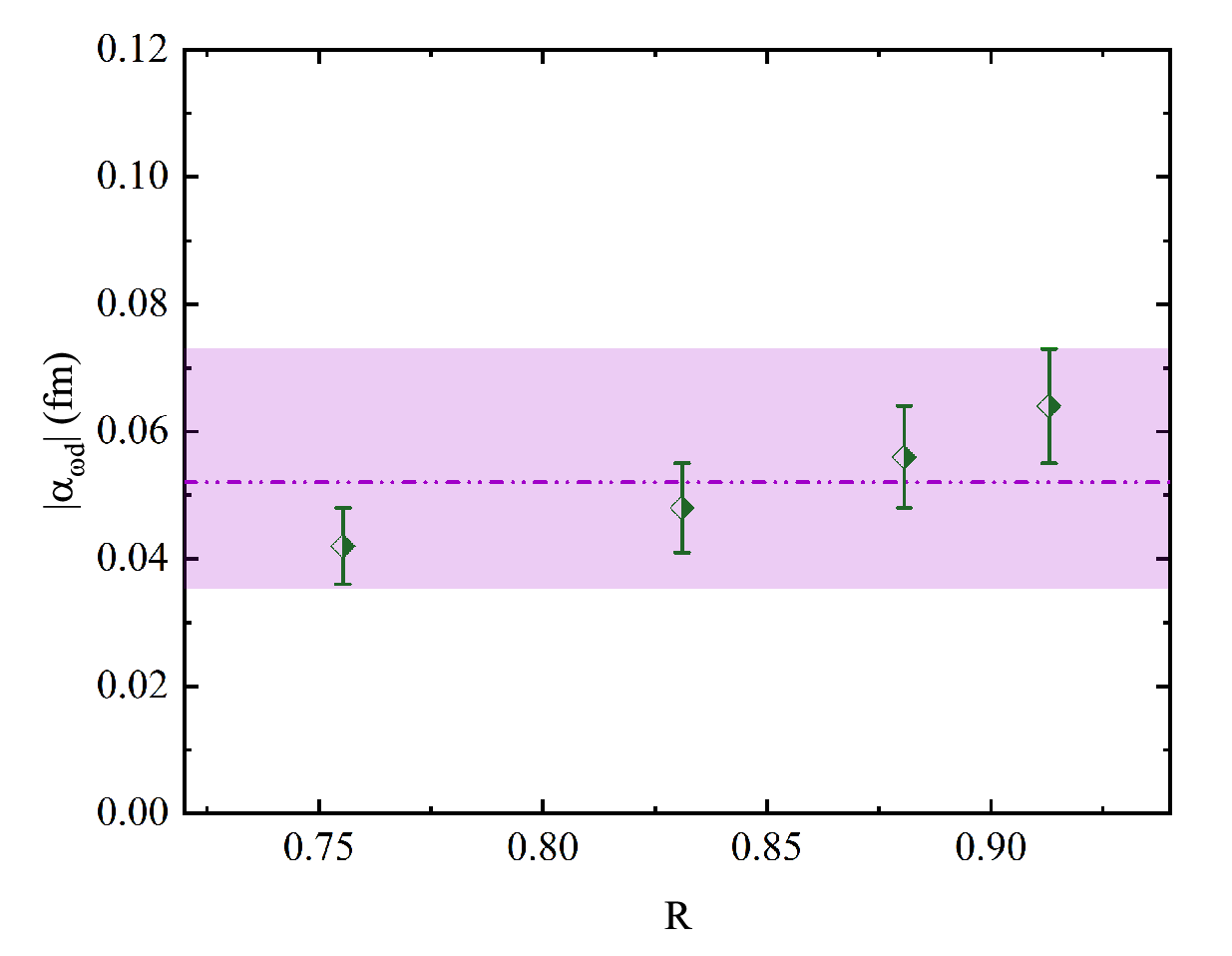}
\end{center}
\caption{The scattering length of $\protect\omega$-$d$ interaction obtained
from the differential cross section of $\protect\omega$ photoproduction off
a deuteron extracted from CLAS at JLAB \protect\cite{CLAS:2018avi}. The
purple double dot dashed line is the RMS value $\protect\sqrt{\left\langle |%
\protect\alpha^{2}_{\protect\omega d}|\right\rangle}=0.052 \pm 0.019$ fm.
The purple band represents the error bar of $|\protect\alpha_{\protect\omega %
d}|$.}
\label{fig:omega-d}
\end{figure}

\begin{table}[htbp]
\caption{The $|\protect\alpha_{\protect\omega d}|$ is obtained from the
different center of mass energies $W$.}
\label{tab:table2}{\small \
\resizebox{\linewidth}{!}{
			\begin{tabular}{ccccc}
				\hline
				\hline
				W (GeV)  & $3.09$ & $3.32$ & $3.59$ & $3.89$ \\
				\hline
				R & $0.76$ & $0.83$ & $0.88$ & $0.91$ \\
				\hline
				$|\alpha_{\omega d}|$ (fm) & $0.041 \pm 0.006$ & $0.048 \pm 0.007$ & $0.055 \pm 0.008$ & $0.064 \pm 0.009$\\
				\hline
				\hline
			\end{tabular}
		}  }
\end{table}

\begin{table}[htbp]
\caption{The experimentally measured the strong interaction and RMS radius
of partial hadrons \protect\cite%
{Povh:1990ad,Solovyev:2020ojc,Klarsfeld:1986lmi}.}
\label{tab:table3}{\small \
\resizebox{\linewidth}{!}{
			\begin{tabular}{ccc}
				\hline
				\hline
				State  & strong interaction radius (fm) & root mean square radius (fm) \\
				\hline
				proton & $0.82\pm0.02$ \cite{Povh:1990ad} & $0.83$ \cite{Solovyev:2020ojc} \\
				
				deuteron& -  & $1.953$ \cite{Klarsfeld:1986lmi}  \\
				
				$\rho$& $0.72\pm0.05$ \cite{Povh:1990ad} &-\\
				
				$\omega$& $0.72\pm0.05$ \cite{Povh:1990ad} & -  \\
				
				$\phi$ & $0.46\pm 0.02$ \cite{Povh:1990ad} & -  \\
				
				$J/\psi$ & $0.20 \pm 0.02$ \cite{Povh:1990ad} & -	\\
				\hline
				\hline	
			\end{tabular}
		}  }
\end{table}

The $J/\psi$ and $\Upsilon$ are also two crucial vector mesons that have
been frequently analyzed. Especially $J/\psi$, as the most basic $c\bar{c}$
state, investigating $J/\psi$-$d$ will lead to the more coherent development
of charmonium. With the fitted function Eq. (\ref{eq:6}), one should predict
the $|\alpha_{J/\psi d}|=0.697 \pm 0.071$ am and $|\alpha_{\Upsilon
d}|=0.260\pm0.084$ am.  Therefore, we can observe the scattering length
distribution of the interaction between vector meson and deuteron. The
scattering length $|\alpha_{Vp}|$ calculated by VMD model is recalled, and
the results combined with $\alpha_{V d}$ are summarized in Fig. \ref%
{fig:comsl}.  There exist a realized phenomenon that the $|\alpha_{V d}|$ is
almost an order of magnitude smaller when compared with the scattering
length of meson-proton $|\alpha_{V p}|$.

The scattering length calculated by the VMD model is supposed as the process
by which vector mesons are embedded into nucleons or nuclei in physics. The
difference in the radius of the hadron involved in the interaction will
influence the depth and degree of embedding.  So these phenomena should be
comprehended in terms of the radius of the hadron. The radius of the strong
interaction and RMS for the different hadron states measured from the
experiment \cite{Povh:1990ad,Solovyev:2020ojc,Klarsfeld:1986lmi} is listed
in Tab. \ref{tab:table3}. In addition, it has been shown that the radius of $\Upsilon$ obtained by non-relativistic potential theory is $0.14$ fm \cite{Satz:2006kba},
 which indicates that the larger the mass of the
vector meson, the smaller the radius.

\begin{figure}[htbp]
	\begin{center}
		\includegraphics[scale=0.40]{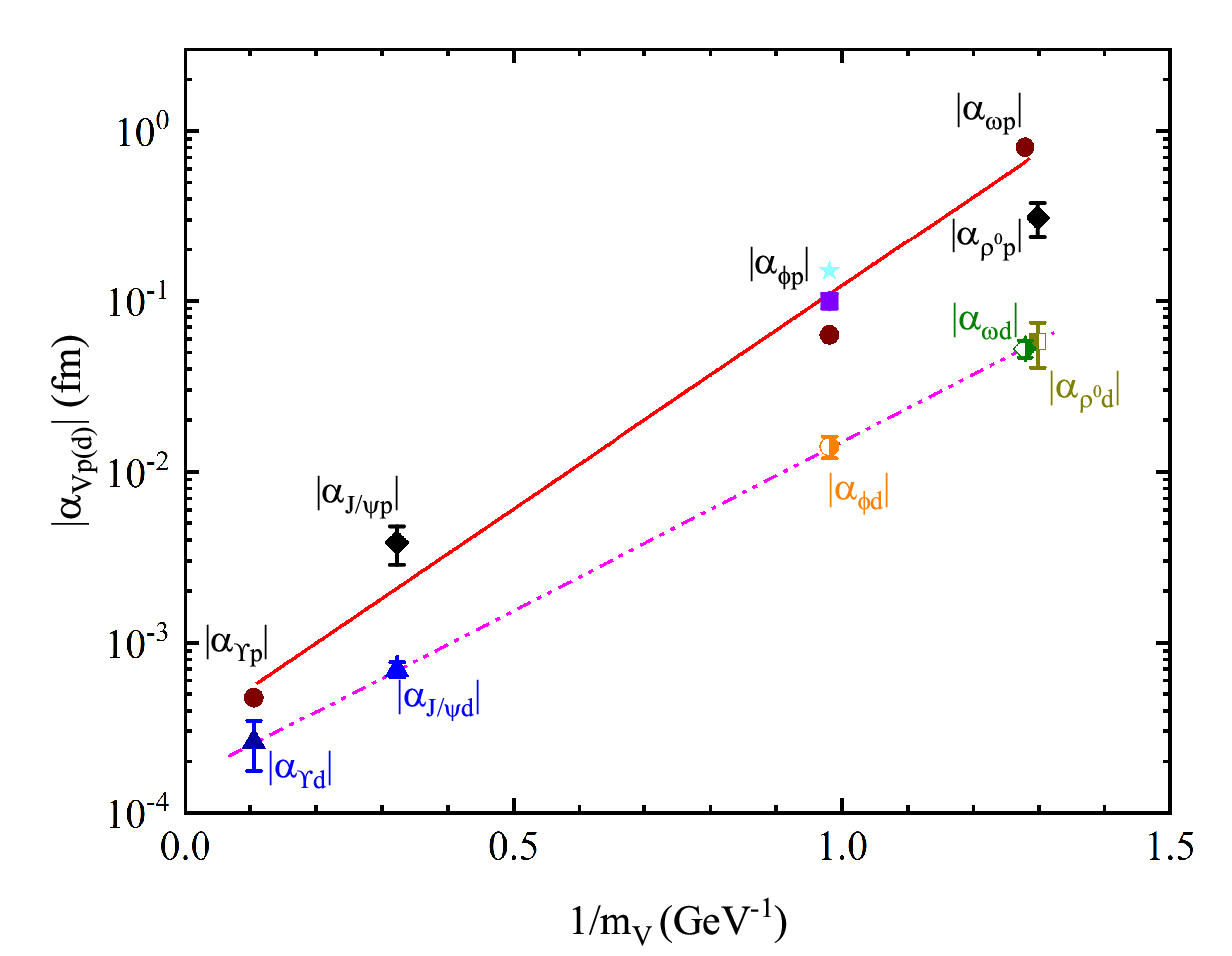}
	\end{center}
	\caption{The scattering length of $V$-$d$ interaction obtained by VMD. The
		burgundy circles for $|\protect\alpha_{\Upsilon p}|$, $|\protect\alpha_{%
			\protect\alpha_{\protect\phi p}}|$ and $|\protect\alpha_{\protect\omega p}|$
		are taken from Igor's work \protect\cite%
		{Strakovsky:2020uqs,Strakovsky:2014wja,Strakovsky:2021vyk}. The result for $|%
		\protect\alpha_{\protect\phi p}|$ represented by the cyan pentagram is
		received from Ref. \protect\cite{Chang:2007fc}. The notations for $|\protect%
		\alpha_{\protect\rho d}|$, $|\protect\alpha_{\protect\omega d}|$ and $|%
		\protect\alpha_{\protect\phi d}|$ are the same as in Fig. \protect\ref%
		{fig:v-d}. The two black diamonds and purple square are derived from our
		previous work \protect\cite{Wang:2022zwz,Wang:2022xpw,Wang:2022tuc}. And the
		two blue triangles are the predicted $|\protect\alpha_{\Upsilon d}|$ and $|%
		\protect\alpha_{J/\protect\psi d}|$.}
	\label{fig:comsl}
\end{figure}

\begin{figure}[t]
	\begin{center}
		\includegraphics[scale=0.60]{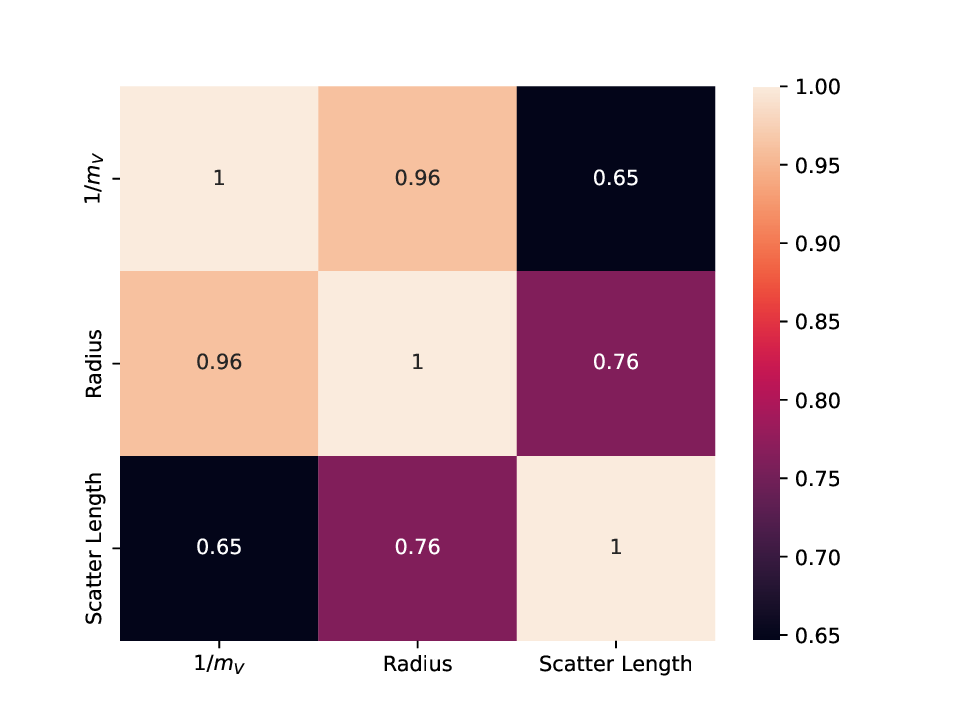}
	\end{center}
	\caption{The strength of the correlation between the three features of $1/m_V$, vector meson radius, and scattering length is indicated by the magnitude of the number. The correlation coefficient closer to $1$ indicates a significantly stronger correlation between the two features.}
	\label{fig:heat}
\end{figure}

The field of machine learning is rapidly expanding and has made its way to high energy physics. This technology offers several benefits to our research, including handling large and complex data sets, extracting features and patterns from incomplete and noisy data, and optimizing and enhancing the performance of physical models and simulations. Unfortunately, without sufficient data, conducting data training and making accurate predictions is not feasible. Therefore, the correlation analysis of scattering length, meson radius and $1/m_V$ is only carried out mathematically.
Using the ``seaborn'' package in Python programming language, we analyzed the Pearson correlation \cite{Parsian M} of these three features,
\begin{equation}
	r=\frac{\sum_{i=1}^n\left(x_i-\bar{X}\right)\left(y_i-\bar{Y}\right)}{\sqrt{\sum_{i=1}^n\left(x_i-\bar{X}\right)^2} \sqrt{\sum_{i=1}^n\left(x_i-\bar{Y}\right)^2}}, \label{eq:7}
\end{equation}
which is shown in Fig. \ref{fig:heat}. In Eq. (\ref{eq:7}), $x_i$ and $y_i$ represent two columns of data features, $\bar{X}$ and $\bar{Y}$ represent the mean of the two columns of data. The correlation coefficient $r$ between two features indicates the strength of their relationship.
When the coefficient approaches $1$, it means that the relationship between the features is stronger. We have observed a positive correlation between the scattering length, radius and $1$/$m_V$. The correlation between the scattering length and the radius is stronger than the correlation between the scattering length and $1$/$m_V$.  The relationship between scattering length and $1$/$m_V$ is clearly demonstrated in Fig. \ref{fig:comsl}, highlighting the significance of the radius of the vector meson in determining the scattering length result. According to the correlation analysis, there is a positive correlation between the radius and $1$/$m_V$. This means that as the mass of the vector meson increases, the radius decreases. Does this mean that the vector meson-proton (deuteron) interaction process can be assumed to be the process of vector meson embedding into proton (deuteron)? The larger the mass and smaller the radius of the vector meson, the deeper it is embedded into the proton (deuteron) and the smaller the scattering length of the interaction.

We attempt to use these three features for further analysis. However, due to the lack of data, the training of data using various algorithms could not produce results with practical significance. Therefore, more theoretical and experimental results are needed to prompt further analysis.

\section{Summary}
\label{sec3}
In this work, the scattering lengths of $\rho$-$d$, $\omega$-$d$ and $\rho$-$%
d$ interaction are investigated under the guidance of the VMD model. Then,
the $|\alpha_{J/\psi d}|$ and $|\alpha_{\Upsilon d}|$ are predicted by a
function which is fitted by the result of $|\alpha_{\omega d}|$ and $%
|\alpha_{\phi d}|$, and the distribution behaviour of the scattering length
of the interaction between vector meson and deuteron is given for the first
time. A few interesting phenomena are discovered by comparing the scattering
lengths of $V$-$p$ and $V$-$d$ interactions.

One finds that the scattering length of vector mesons interacting with protons or deuterons is proportional to a specific portion of the mass of vector mesons, with the exception of the scatter length of $\rho$-$p$/$d$. Using the correlation analysis method in mathematics, we analyzed the correlation between $1$/$m_V$, radius of vector meson, and scattering length. Our findings suggest a strong positive correlation between scattering length and both $1$/$m_V$ and radius of vector meson, with a high correlation coefficient. In other words, the larger the radius of the vector meson, the greater the scattering length generated by its interaction with the proton or deuteron. From this, we assume that the scattering length is related to the degree to which vector mesons are embedded in protons or deuterons. Unfortunately, the lack of available data makes it difficult to proceed with machine learning for the next training phase.

Furthermore, considering the assumption that the scattering length between heavy vector mesons and nucleus is smaller than that between light vector mesons and nucleus, does this may imply that the embedding degree between heavy vector mesons and nucleus is higher, making it easier to form $J/\psi N$ or $%
\Upsilon N$ combination of pentaquark states \cite%
{LHCb:2015yax,LHCb:2019kea,Wang:2019krd,Wang:2019zaw} or molecular states?
Therefore, we suggest that experiments can strengthen research in this area,
such as studying the production of pentaquark states in the heavy quark
energy region through hadron-hadron scattering \cite{Wang:2019dsi}, which can effectively avoid the influence of kinematic singularities in the process of pentaquark states production through heavy hadron decay. Our study on the scattering length between vector mesons and nucleus may provide useful theoretical references for better understanding the production of pentaquark states or the binding energy in molecular states \cite%
{Kaidalov:1992hd}.

Overall, this work may provide an important insights into the interaction of
vector mesons with nucleons or nucleus. However, the lack of experimental
data limits our ability to draw definitive conclusions, and additional
experimental data on the $\gamma d$$\rightarrow$$V d$ scattering cross
section is needed to further refine our understanding of this process.

\begin{acknowledgments}
		This work is supported by the National Natural Science Foundation of China under Grants No. 12065014 and No. 12047501, and by the Natural Science Foundation of Gansu province under Grant No. 22JR5RA266. We acknowledge the West Light Foundation of The Chinese Academy of Sciences, Grant No. 21JR7RA201.
\end{acknowledgments}

\end{document}